\pdfoutput=1
\documentclass{du-journals}
\journal{Iranian Journal of Astronomy and Astrophysics}
\title{The role of sunspots magnetic configuration in the formation of umbral fine-structures\footnote{Published in IJAA, 2023, Vol. 10, Issue 1, Pages 111-121, DOI: 10.22128/IJAA.2023.717.1161 (https://ijaa.du.ac.ir/).}}
\author{H. Hamedivafa}
\address{Physics Department, Faculty of Science, Imam Khomeini International University,
Qazvin 34149-16818, Islamic Republic of Iran; email:
vafa@sci.ikiu.ac.ir}
\begin{document}
\begin{abstract}
We use spectro-polarimetric data recorded by Hinode to analyze the
magnetic field configuration of a part of a sunspot (AR~10923) where
a bundle of penumbral filaments are intruding into its umbra. We
want to explore the role of the sunspot magnetic configuration in
the formation and kinematics of the fine-structures, such as umbral
dots and light bridges, inside the sunspot umbra. Both direct
inferences from polarization Stokes profiles and the inversion
results using the SIR code indicate an aligned magnetic field
configuration in the umbra where moving umbral dots are easily
formed at the leading edges of the rapidly intruding penumbral
filaments. We suggest that the magnetic field topology is rearranged
leading to the observed aligned magnetic field lines via magnetic
reconnection process by which a part of the magnetic energy is
converted into thermal and kinetic energy. This new configuration
causes the umbral fine-structures to form easily and more frequent.
\end{abstract}

\begin{keywords}
  Sun: sunspots, fine-structures, magnetic configuration
\end{keywords}

\section{Introduction}\label{intro}
Sunspots, the darkest regions on the solar surface, are understood
to be the result of partially inhibiting heat transfer from the
solar interior to the surface by strong magnetic fields
\cite{sol03,remp09}. The investigation of the bright fine-structures
formed inside sunspot umbrae, like umbral dots or light bridges
(LBs), are important to understand the physical mechanisms
responsible for the heat transfer into the photosphere and for the
magnetic field energy dissipation in atmospheric layers. LBs are
lanes of relatively bright and hot plasma dividing sunspot umbrae
into two parts and displaying a large variety of morphology and fine
structure \cite{sob97asp} and can be seen during both formation and
decaying of a sunspot (for a review see \cite{morad10}).

The origin of dark cores/lanes observed in bright penumbral
filaments \cite{scharm02} as well as in LBs in continuum and
polarization maps was theoretically described by Ruiz Cobo \& Bellot
Rubio \cite{cobo08}: a dark core is a consequence of the higher
density of the plasma inside penumbral filament, which shifts the
surface of optical depth unity toward higher (cooler) layers.

It is still unknown what triggers the formation of an LB, and what
is the role of the penumbra in the development of an LB
\cite{kats07} and in the formation of umbral dots \cite{kats07,sv06}
and their kinematics \cite{sob97aa,wata10}.

In this paper, the umbral magnetic structure around a bundle of
penumbral filaments intruding into a sunspot umbra is investigated
using spectro-polarimetric measurements from the Japanese satellite
Hinode \cite{koso07}. In the studied umbral area, a long and faint
filamentary LB (as classified by Sobotka \cite{sob97asp}) is seen.

\section{Data Set}\label{dataset}
The data analyzed here were obtained using the spectro-polarimeter
of the Solar Optical Telescope (SOT: \cite{tsu08, suem08}) onboard
Hinode satellite. The instrument observes the two iron lines Fe~I
630.15~nm ($g=1.67$) and Fe~I 630.25~nm ($g=2.5$), whose line
formation region is between $\log(\tau) \approx 0$ and $-3$.  A
normal full-field scan of AR~10923 was taken on 15 November 2006,
during the period 03:15 - 03:52 UT, when the spot was located at the
heliocentric angle of $13^\circ$. From this scan, a field of
$20\times20$~arcsec~$^{2}$, corresponding to 130 consecutive slit
positions and containing the part of the penumbra intruding into the
umbra, was selected for further analysis. The continuum intensities
of the Stokes~\textit{I} profiles from the pixels with low
polarization signals in the whole field of view and the known limb
darkening function were used to find the continuum intensity of the
quiet Sun at disk center to normalize all Stokes profiles. The
sunspot has negative magnetic polarity, and then the regular
Stokes~\textit{V} profiles have negative blue lobes.

\begin{figure} %Fig.~1
 \centerline{\includegraphics[width=0.4\textwidth]{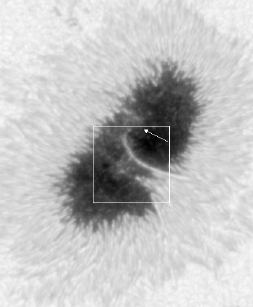}}
 \caption[]{The continuum intensity map reconstructed from the continuum of the
 Stokes~\textit{I} profiles. The size of the image is $430\times520$~pixels,
 equivalent to $68\times82$~arcsec$^{2}$. The size of analyzed region (white
 square) is $130\times130$~pixels, equivalent to $20\times20$~arcsec$^{2}$.
 The arrow points to solar disk center.}
  \label{conti}
\end{figure}

Figure~\ref{conti} shows the intensity map of the sunspot as
reconstructed from the continuum intensities of the
Stokes~\textit{I} profiles. The area ($130\times130$~pixels)
analyzed here is shown by a white square in Figure~\ref{conti}. This
area contains a long filamentary light bridge (LB), a part of the
penumbra intruding into the umbra and some individual umbral dots
(UDs) and chains of them. The formation process of this LB was
studied by Katsukawa~et~al. \cite{kats07}. They suggested that the
continual emergence of umbral dots from the leading edges of
penumbral filaments as well as their rapid inward migration forms
the LB inside the umbra.

%----------------------------------------------------------------------
%----------------------------------------------------------------------

\section{Direct Inferences from Observed Stokes Profiles}\label{direct}
\subsection{Fine structures of the filamentary light bridge}\label{fs_lb}

As observed with Hinode \cite{rubio07} and the SST \cite{noort08},
dark cores is more prominent in polarized light (Stokes~\textit{Q,
U} and~\textit{V}) than in continuum intensity. However, Bellot
Rubio~et~al. \cite{rubio07} found weaker dark core polarization
signals. Their inversion results indicate that dark cores have
weaker and more inclined magnetic fields.

\begin{figure} %Fig.~2
 \centerline{\includegraphics[width=0.9\textwidth,clip=]{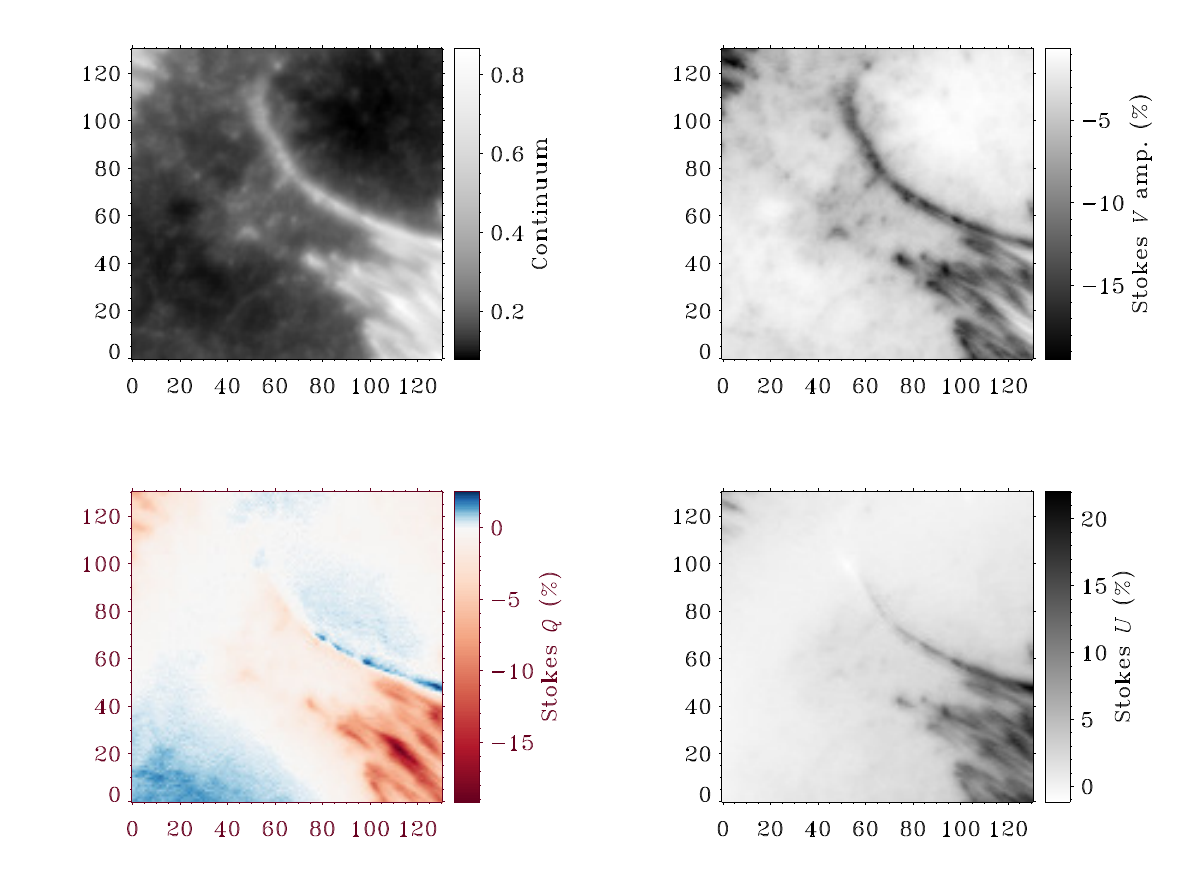}}
 \caption[]{Left to right, top to bottom: continuum intensity map, the map of
 Stokes~\textit{V} amplitude, and the maps of \textit{Q} and \textit{U} signals at
 the line core. All maps are reconstructed from the 630.25~nm line and normalized
 to the quiet Sun continuum intensity at disc center.
}
 \label{polmaps}
\end{figure}

\begin{figure} %Fig.~ 3
 \centerline{\includegraphics[width=0.9\textwidth,clip=]{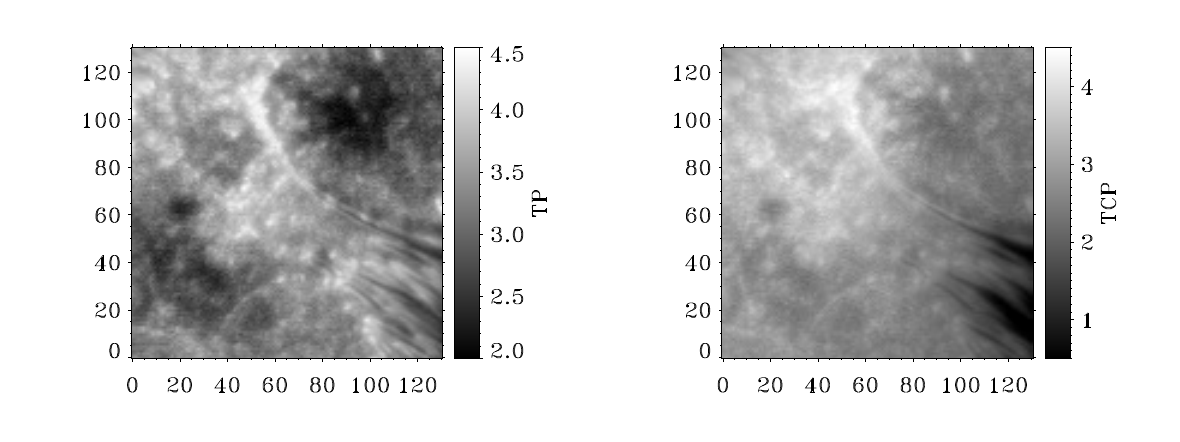}}
 \caption[]{Left panel: total polarization map. Right panel: total circular
 polarization map. All maps are reconstructed from the 630.25~nm line and
 normalized to the quiet Sun continuum intensity at disc center.
 }
 \label{tptcp}
\end{figure}

Figure~\ref{polmaps} shows the continuum intensity map as well as
the maps of Stokes~\textit{Q, U} and~\textit{V} computed from
630.25~nm line. Here \textit{Q} and \textit{U} refer to the signal
values of the $\pi$ component of the corresponding Stokes parameters
(signal at the line core), and \textit{V} refers to the (negative)
amplitude of the blue lobe of Stokes \textit{V} profile. We cannot
clearly resolve any dark core on the filamentary LB in the continuum
intensity map \cite{cobo08} (see upper-left panel in
Figure~\ref{polmaps}). Figure~\ref{tptcp} shows total polarization
(TP~$=\sum_{\lambda}{\sqrt{Q^{2}+U^{2}+V^{2}}}$), left panel) and
total circular polarization (TCP~$=\sum_{\lambda}{|V|}$), right
panel) maps, both constructed from the 630.25~nm line. Although
spatial resolution of Hinode/SP observations is not quit high,
however, in some Stokes polarization maps, we can clearly resolve
dark core on the filamentary LB as well as on other intruding
filaments at the lower-right corner of the studied area
\cite{rubio07, noort08}: stronger signals in Stokes \textit{V} and
\textit{U} (upper- and lower-right panels in Figure~\ref{polmaps},
respectively), weaker signals in TP and TCP maps (left and right
panels in Figure~\ref{tptcp}, respectively). However, as can be seen
in the lower-left panel in Figure~\ref{polmaps}, the dark core in
the map of Stokes~\textit{Q} signals has a different display: The
core of the LB appears as a sharp border dividing positive and
negative signals.

%----------------------------------------------------------------------
%----------------------------------------------------------------------

%----------------------------------------------------------------------
%----------------------------------------------------------------------

\subsection{Other fine structures}\label{fs}
All bright structures especially umbral dots resolved in the map of
continuum intensity (upper-left panel in Figure~\ref{polmaps}) are
more pronounced in TP map (left panel in Figure~\ref{tptcp}). A
faint tinny arced structure seen at the lower half of the continuum
intensity map can also be recognized in the maps of
Stokes~\textit{V} (Figure~\ref{polmaps}) as well as in the maps of
TP and TCP (Figure~\ref{tptcp}). Also a few chains of small umbral
dots are seen at the center of the maps of continuum intensity,
Stokes~\textit{V} signals and TP.

As mentioned before, the maps displayed in Figures~\ref{polmaps} and
\ref{tptcp} were computed from the 630.25~nm line. The corresponding
maps derived from the 630.15~nm line are very similar to the ones
reconstructed from the 630.25~nm line. We concentrate on the results
of the 630.25~nm line which is a triplet Zeeman line.

%----------------------------------------------------------------------
\begin{figure} %Fig.~4
 \centerline{\includegraphics[width=0.9\textwidth,clip=]{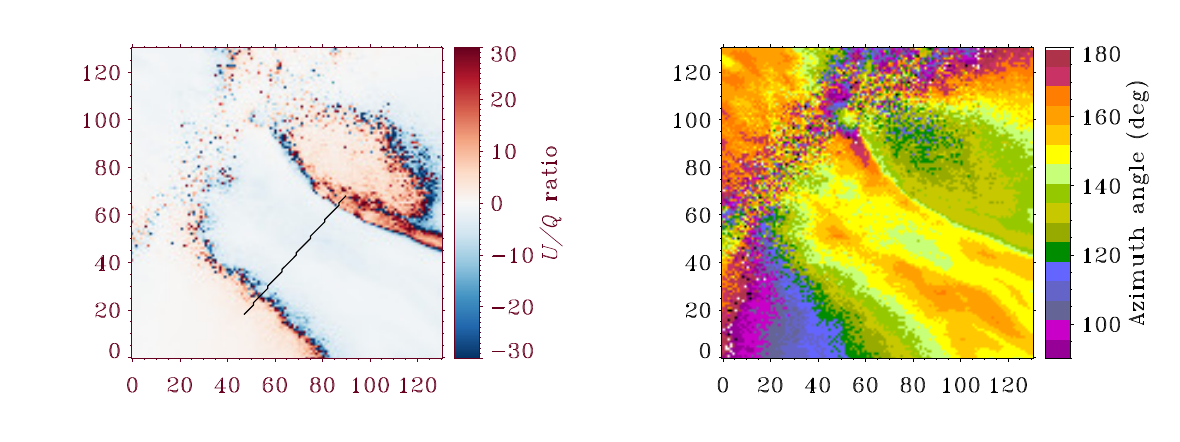}}
 \caption[]{Left panel: the map of \textit{U/Q} ratio. The dashed line shows a
 cut across the region whose polarization signals are displayed in Figure~\ref{acrosscut}.
 Right panel: the map of the azimuth angle of the magnetic field vector estimated from
  $\tan(2\varphi)=U/Q$. We did not correct the $180^\circ$ ambiguity in azimuth angle
  in the upper-left corner of the studied area.
 }
 \label{uqratio}
\end{figure}
%----------------------------------------------------------------------

\subsection{Azimuth angle of the magnetic field}\label{fieldazimuth}
The map of Stokes~\textit{Q} signals shown in the lower-left panel
in Figure~\ref{polmaps} needs special attention. This map shows a
distinct (red) region having negative values extended from the
center of the map to the lower-right corner including the bright
structures (penumbral grains and filaments) seen in the continuum
intensity map. This may be questioned that this behavior is the
natural consequence of the measured Stokes~\textit{Q} or
\textit{U}on a fanning magnetic structure of a sunspot in Hinode/SP
observations. However, both lower and upper seen boundaries cannot
be explained. To investigate what happened in this region we compute
the ratio of \textit{U/Q} which scales with $\tan(2\varphi)$ in
which $\varphi$ is the azimuth angle of the magnetic field vector
\cite{auer77, sten10}. The left panel in Figure~\ref{uqratio}
displays the map of this ratio in which the distinct region with a
clearly sharp boundary is seen as it was observed in the map of
Stokes~\textit{Q} signals (lower-left panel in
Figure~\ref{polmaps}). The upper boundary of the region is
co-spatial with the filamentary LB. The map of the azimuth angle of
the magnetic field vector estimated using $\tan(2\varphi)=U/Q$ is
shown in the right panel in Figure~\ref{uqratio}. We do not resolve
the $180^\circ$ azimuth ambiguity in the upper-left corner where is
not area of interest.

%----------------------------------------------------------------------
\begin{figure} %Fig.~5
 \centerline{\includegraphics[width=0.9\textwidth,clip=]{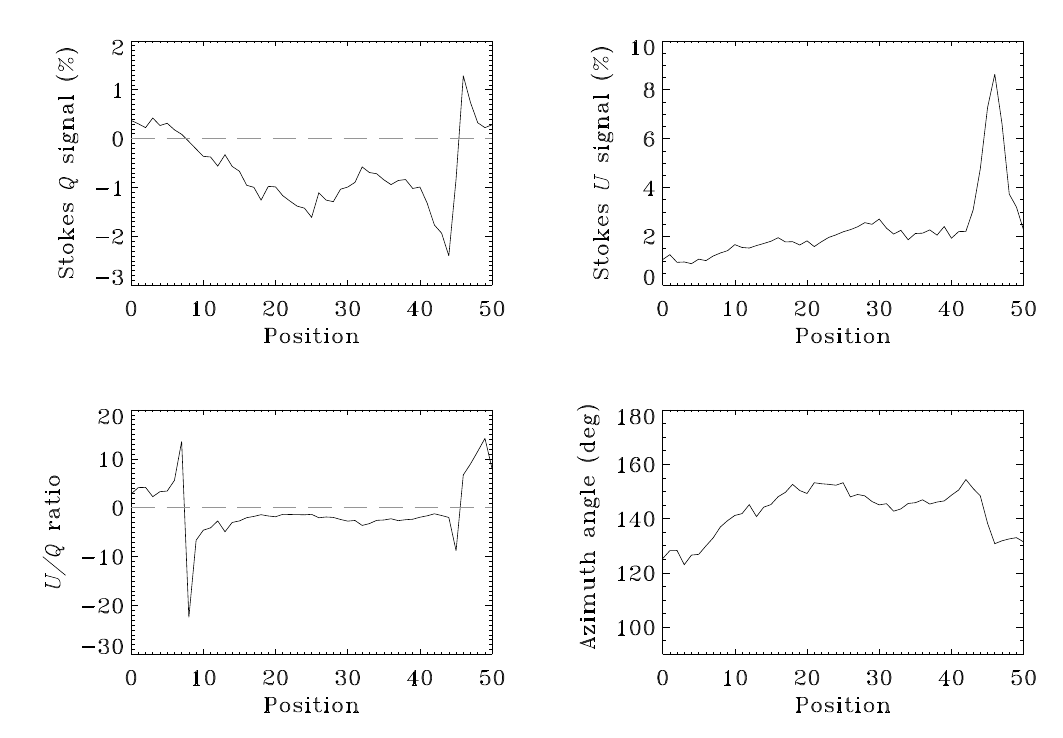}}
 \caption[]{Line core signals of Stokes~\textit{Q} and \textit{U} (upper panels),
 and the values of \textit{U/Q} ratio (lower-left panel) along the cut shown by
 dashed-line in Figure~\ref{uqratio}. Lower-right panel: the plot of the azimuth
 angle ($\varphi$) of the magnetic field along the cut, estimated from $\tan(2\varphi)=U/Q$.
 }
 \label{acrosscut}
\end{figure}
%----------------------------------------------------------------------

Figure~\ref{acrosscut} shows the variations of the signals of
Stokes~\textit{Q} and \textit{U} (upper-left and -right panels,
respectively), and values of the \textit{U/Q} ratio (lower-left
panel) along the shown cut across the distinct region (see
Figure~\ref{uqratio}). The magnetic field azimuth angles along the
cut, estimated using $\tan(2\varphi)=U/Q$ are also shown in the
lower-right panel in Figure~\ref{acrosscut}. The signals of
Stokes~\textit{Q} (upper-left panel) are positive outside the
distinct region, decreasing to zero where borders are defined.
Entering the distinct region, the signals of Stokes~\textit{Q} show
negative non-zero values. Stokes~\textit{U} signals (upper-right
panel) are positive inside and outside of the distinct region. The
different values of the \textit{U/Q} ratio outside and inside the
distinct region define different azimuth angles for the magnetic
field vectors around the border of the distinct region.

\begin{figure} %Fig.~6
 \centerline{\includegraphics[width=\textwidth,clip=]{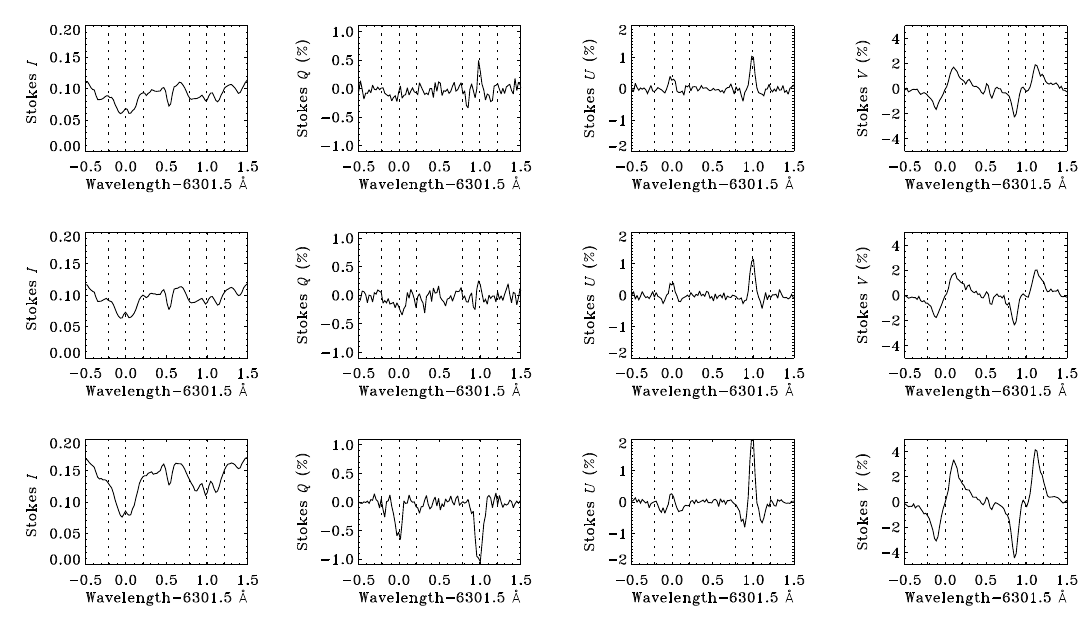}}
 \caption[]{The profiles of Stokes~\textit{I, Q, U} and~\textit{V} for three
 selected pixels on the cut shown by dashed-line in Figure~\ref{uqratio} at
 intersecting the lower border of the distinct region. Upper (lower) panels:
 Stokes profiles for a pixel outside (inside) the distinct region. Middle panels:
 Stokes profiles for a pixel on the border. The vertical dashed lines show the
 line core and $\pm215$~m\AA~around it in both spectral lines.
 }
 \label{stksprofs}
\end{figure}

Figure~\ref{stksprofs} displays the full Stokes profiles for three
pixels around the lower boundary of the distinct region on the
defined cut. We can see the changes of the signal at the line core
of Stokes~\textit{Q} when passing the boundary.

Looking at the right panel in Figure~\ref{uqratio}, we can see that
the distinct region show an area where the magnetic field vectors
have almost the same azimuth angles implying \textit{aligned}
magnetic field vectors. This means that the magnetic field vector
noticeably rotates by changing its azimuth upon entering the
distinct region. This can be an evidence for a process rearranging
the magnetic field vectors in the distinct region. In other words,
the magnetic field vectors tend to be well-arranged there.

%----------------------------------------------------------------------
%----------------------------------------------------------------------

\section{Inversion Results}\label{inversion}
To have a confirmation for the direct inferences we applied the
inversion code SIR (Stokes Inversion based on Response functions
\cite{cobo92}) to the observed spectra. Inversions were made for a
single-component magnetic model atmosphere. As discussed by Sobotka
\& Jur$\check{c}\acute{a}$k \cite{sob09}, the instrumental scattered
light is negligible in Hinode/SP data.

Blending spectral lines are more prominent in Stokes profiles
observed in a sunspot umbra (see the left panels in
Figure~\ref{stksprofs}). As shown by Hamedivafa \cite{vafa13} the
unknown blending lines broaden the recovered Stokes~\textit{I} cause
a higher macro/microturbulent velocity and give smaller
near-continuum intensity, resulting in slightly cooler temperature
stratification at least in deeper layers. Therefore, the code
ignores the blended parts of the spectral lines (see Figure~9 in
\cite{vafa13}) to recover a better fitted profile and to retrieve a
more reliable atmospheric model.

To detect the changes of temperature with height, we used five nodes
for temperature and two nodes for LOS velocity and magnetic field
strength in the final step of the inversion process. Furthermore, to
find effective values for the magnetic field inclination and its
azimuth angle, we considered them height-independent (one node).
Additionally, we enabled one node for micro-turbulent velocity. The
weights of Stokes~\textit{I, Q, U} and~\textit{V} were adopted to be
equal in the inversion.

\begin{figure} %Fig.~7
 \centerline{\includegraphics[width=\textwidth,clip=]{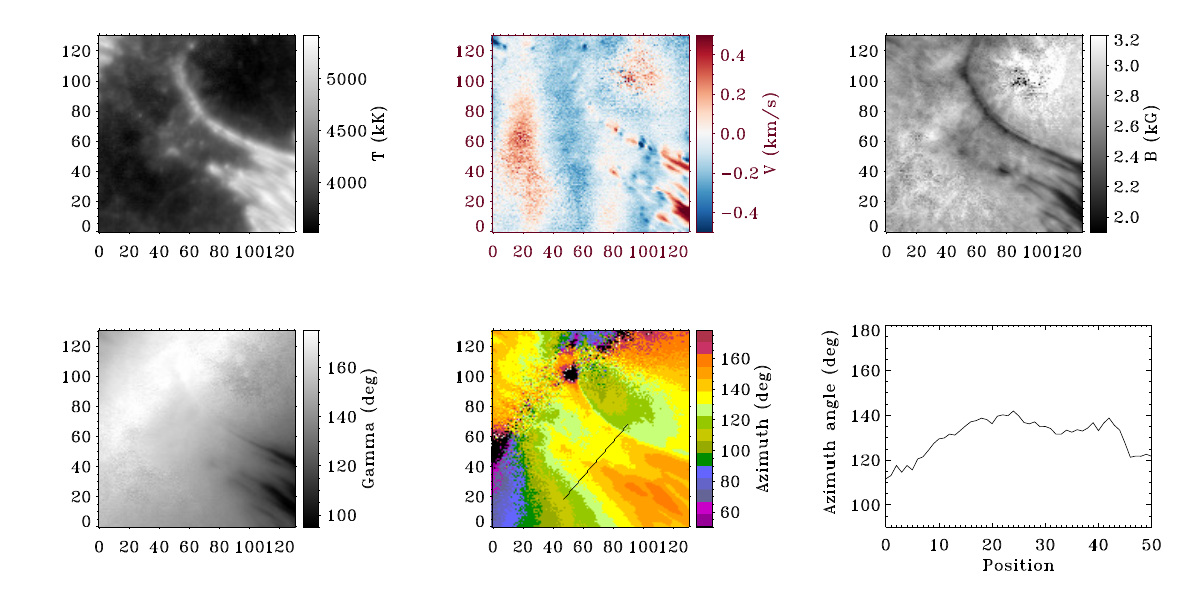}}
 \caption[]{Upper panels, left to right: maps of temperature (T), magnetic field
 strength (B), LOS velocity (V). Negative velocities (blue) are upflows. Lower-left
 and -middle panels: maps of inclination (Gamma) and azimuth angle at optical depth
 of $-0.4$. Lower-right panel: the graph of azimuth angles derived from the retrieved
 atmospheric model across the distinct region along the cut shown in the azimuth panel.
 }
 \label{atmparam}
\end{figure}

Figure~\ref{atmparam} displays the retrieved photospheric model at
optical depth of $-0.4$. Although the temperature (T) map is similar
to the intensity map shown in Figure~\ref{polmaps}, however, in the
temperature (T) map we can recognize dark cores on the penumbral
filaments with a good contrast. Velocities have been calibrated
considering zero velocity for the umbral core region. In the
velocity map (V), adjacently localized small patches showing up- and
down-flows are seen on the LB implying a multi-segmented structure
for it \cite{thomas04}. This structure is similar to the multi-cell
convection pattern for light bridges suggested by
Jingwen~Zhang~et~al. \cite{jing18}. Penumbral grains, the bright
head of the intruding penumbral filaments, show upflows up to
400~m~s$^{-1}$, although the signature of the p-mode oscillations
enters a perturbation in the velocity map. Also, at the lower-right
side of the studied area, the LOS component of the outward Evershed
flow \cite{sol03} is seen as downflows along the intruding penumbral
filaments. Magnetic field on the core of bright filamentary
structures and penumbral grains is weaker (B map) and more inclined
(Gamma map) than that in their surroundings. This view is consistent
with the inversion results of Bellot Rubio~et~al. \cite{rubio07}.
The curved filamentary LB has a signature in the map of azimuth
angle as well. Here, also we did not correct the $180^\circ$ azimuth
ambiguity in the upper-left corner of the studied area. The distinct
region shows a uniform azimuth angle but different (in average,
$15^\circ$ larger) with the azimuth of the surrounding filed.

The variations of the azimuth angle across the distinct region along
the defined cut shown in lower-middle panel in Figure~\ref{atmparam}
(the same as shown in Figure~\ref{uqratio}) are plotted in the
lower-right panel in Figure~\ref{atmparam}. These variations are
very similar to the corresponding variations plotted in the
lower-right panel in Figure~\ref{acrosscut}, although the values
have a difference of about $10^\circ$, in average. The difference
arises from the different ways obtaining the azimuth. In direct
inference we used the amplitudes of Stokes \textit{Q} and \textit{U}
at the line center and in inversion we adopted a height-independent
azimuth. The inversion results confirm the direct inferences for
azimuth angle shown in Figures~\ref{uqratio} and \ref{acrosscut}.
Also, the distinct region shows itself in the inclination map
(Gamma) as a region with an equalized inclination angle.

%----------------------------------------------------------------------
%----------------------------------------------------------------------

%----------------------------------------------------------------------
\begin{figure} %Fig.~8
 \centerline{\includegraphics[width=0.7\textwidth,clip=]{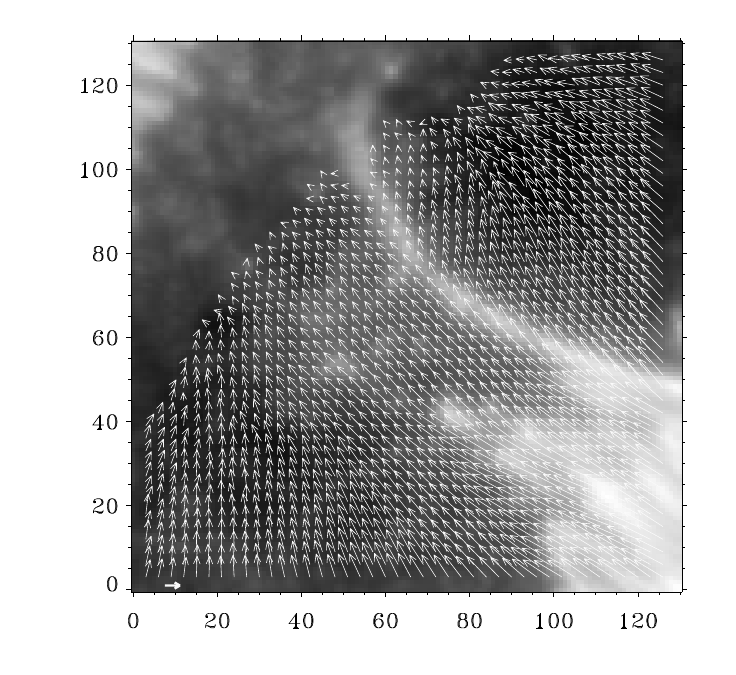}}
 \caption[]{The projection of the magnetic field vectors in the plane perpendicular
  to the LOS. Horizontal vector at the lower-left corner of the map displays
  the unit vector (1~kG).
 }
 \label{vectfield}
\end{figure}
%----------------------------------------------------------------------

Figure~\ref{vectfield} shows the projection of the magnetic field
vector in the plane perpendicular to the LOS. Since the sunspot is
close to the solar disc center, it does not need the magnetic field
vectors to be transferred from the LOS reference frame into the
local (disc center) reference frame. Inside the distinct area, field
vectors are (\textit{orderly}) \textit{aligned} as we deduced in
Section~\ref{fieldazimuth}.

\section{Summary and Conclusion}\label{summary}

Here we presented an investigation of SP observations recorded by
Hinode/SOT from a part of sunspot umbrae (AR~10923). The studied
area contains a long filamentary light bridge (LB), penumbral
filaments intruding into the umbrae and umbral dots. The spot was
located close to the disk center. Our investigation consists of the
results of both direct analysis of Stokes spectra and inversion
using SIR code. In this study, we explore the role of the penumbra
in the formation and kinematics of umbral dots and LBs.

Katsukawa~et~al. \cite{kats07} studied the formation process of the
LB observed in our studied area. They pointed out that during the
formation of the LB, many umbral dots were observed, emerging from
the leading edges of the rapidly intruding penumbral filaments. They
identified central umbral dots formed inside the umbra with a
relatively slow inward motion as the precursor of the LB formation.
But, what process causes mobile umbral dots to be easily formed
inside the umbra. They suggested that the emergence and the inward
motion of the central umbral dots are triggered by a buoyant
penumbral flux tube as well as upward subphotospheric flows reaching
the surface.

Our study shows that umbral dots and intruding penumbral filaments
reside in a distinct region where the magnetic field vectors are
rearranged so that their azimuths as well as inclinations are
equalized. The LB forms one boundary of the distinct region. We
suggest the orderly rearranging of the magnetic field vectors is the
reason of the easily formation of the umbral dots and their slow
inward motion observed by Katsukawa~et~al. \cite{kats07}.

Chromospheric observations reveal long-lasting recurring jet-like
activities above light bridges (\textit{e.g.}, \cite{roy73, song17})
or inverted Y-shaped magnetic structures in a penumbral region
\cite{zeng16} and a penumbral intrusion into the umbra \cite{bha17}.
These are usually suggested to be a natural consequence of magnetic
reconnection by which a part of the magnetic energy is converted
into thermal and kinetic energy (\textit{e.g.}, \cite{zhang12}).
Therefore, the magnetic field topology is rearranged via magnetic
reconnection process. This rearrangement may lead to a well-arranged
magnetic field configuration. Therefore, it is also important to
study the photospheric and chromospheric magnetic configuration of a
sunspot before intruding penumbral filaments into the umbra to
confirm this suggestion.

\section*{Acknowledgment}
Hinode is a Japanese mission developed and launched by ISAS/JAXA,
with NAOJ as domestic partner and NASA and UKSA as international
partners. It is operated by these agencies in cooperation with ESA
and NSC (Norway). Data analysis was, in part, carried out on the
Multi-wavelength Data Analysis System (MDAS) operated by the
Astronomy Data Center (ADC), National Astronomical Observatory of
Japan. The author sincerely thanks Yukio Katsukawa for his supports
for MDAS user account.

\end{document}